\documentclass{jpsj2}
\newcommand{\lan }{\langle}
\newcommand{\ran }{\rangle}
\newcommand{\lr }{\left(}
\newcommand{\rl }{\right)}
\newcommand{\beq}{\begin{eqnarray}}
\newcommand{\eeq}{\end{eqnarray}}
\newcommand{\ii }{{\rm i}}
\newcommand{\ee }{{\rm e}}

\title{Asymmetric Heat Flow in Mesoscopic Magnetic System}

\author{\textsc{Keiji Saito}}

\inst{Department of Physics, Graduate School of Science,
University of Tokyo, Hongo 7-3-1, Bunkyo-ku, Tokyo 113-0033, Japan
}

\abst{
The characteristics of heat flow in a coupled magnetic system are studied.
The coupled system is composed of a gapped chain and a gapless chain.
The system size is assumed to be quite small so that the mean free path 
is comparable to it. 
When the parameter set of the temperatures of reservoirs is exchanged,
the characteristics of heat flow are studied with the Keldysh 
Green function technique.
The asymmetry of current is found in the presence of
a local equilibrium process at the contact between 
the magnetic systems. 
The present setup is realistic and such an effect will be observed in real 
experiments.
We also discuss the simple phenomenological explanation to
obtain the asymmetry. 
}

\kword{quantum thermal conduction, heat flow control, quantum magnetic systems}

\begin{document}
\maketitle

\section{Introduction} 
Recently, quantum heat transport in magnetic materials has attracted much 
interest \cite{SFGOVR00,SFGOVR00b,ATSDTFK98,HLUKZFKU01,SGOARBT00,SGOAR00,KINAKMTK01}. 
In many magnetic materials, a magnetic contribution 
is dominant in heat transport in comparison with phononic heat transport.
Generally the thermal conductivity is sensitive to magnetic phase 
transitions. Therefore, measurements of heat flow provide information
on magnetic properties \cite{ATSDTFK98}. On the other hand, 
magnetic materials 
have also provided fundamental information on heat transport in nonequilibrium 
statistical mechanics. For example, which types of 
magnetic system show a diffusive transport in 
the thermodynamic limit?
In an isotropic Heisenberg chain, the heat flow 
is a conserved quantity, and
ballistic heat transport is theoretically expected 
\cite{ZNP97}. Generally integrable systems show a divergent conductivity
\cite{ZNP97,STM96}.
In Sr$_{2}$CuO$_{3}$ and CuGeO$_{3}$, which are well described 
by the isotropic Heisenberg chain,
the ballistic feature of heat transport is actually confirmed by 
measuring an unusually large mean free path ($\sim$ 500-1000 times the 
lattice constant)
\cite{SFGOVR00,SFGOVR00b,ATSDTFK98}. 
Interestingly, the Heisenberg chain with extra perturbation terms also 
shows very large conductivity, which is shown for many materials such as
the spin-Ladder systems (Sr,La,Ca)$_{14}$Cu$_{24}$O$_{41}$ and 
La$_{5}$Ca$_{9}$Cu$_{24}$O$_{41}$\cite{SGOARBT00,SGOAR00,KINAKMTK01}.
Although in real materials, macroscopic heat transport in the 
thermodynamic limit cannot be ballistic because of the existence of
spin-phonon interaction, impurities and so on, it is still of academic 
interest to study the thermodynamic behavior of magnetic heat transport.
Therefore, both theoretically and experimentally, 
the properties of energy transport 
in the thermodynamic limit have been intensively studied 
\cite{theo1,theo2,theo3,theo4,theo5,theo6,theo7,theo8,HHCB02}.
 
In this paper, however, we do not study the thermodynamic limit. We focus on 
mesoscopic-scale systems, where mean free path is comparable to 
system size. Therefore, phononic heat transport is negligible
at very low temperatures, and magnetic heat transport is regarded as
ballistic.
We study a coupled magnetic systems and consider how the heat flow can 
be controlled. In this case, two magnetic materials are connected to each other, and 
it is assumed that one magnetic system is gapless, and the other 
is gapped.
In this case, there exists no magnetic interaction between the 
two magnetic chains.
Therefore, the heat is conveyed through phonons at the surface of contact 
between them.
Here, we ask what should be expected for heat flow in this setup. 

A similar setup has been studied in classical anharmonic chains, where
left and right chains have different phononic mode distributions. 
At the left and right edges, thermal reservoirs at different temperatures 
are attached.
Terraneo et al. numerically showed that asymmetric heat flow is 
observed when the temperatures at both edges are exchanged \cite{TPC02,LWC04}. 
This phenomenon is observed only for the regime far from equilibrium, 
and cannot be explained within the linear response theorem.
Recently, it has been desired to demonstrate this phenomenon in real 
experiments \cite{SN}.
In this paper, we propose a promising experimental setup 
to observe asymmetric heat flow using the above coupled magnetic system. 
It is emphasized that our setup 
is based on a {\em mesoscopic quantum magnetic system} so that 
we have a wide choice of magnetic material.

Here, we theoretically analyze heat flow using the bosonization 
technique \cite{text}.
The bosonized Hamiltonian is convenient because it guarantees ballistic 
energy flow, which is expected in a mesoscopic regime \cite{S03}.
We adopt the Keldysh Green function technique for the boson operators
to study the nonequilibrium stationary state.
We show that asymmetric heat flow is actually observed only when the 
thermalization of phonons occurs, which is normally expected in real 
experiments.

This paper is organized as follows. In \S $2$, we introduce the 
model, and analyze the stationary state using the Keldysh Green function 
method in \S $3$.
In \S $4$, another analysis to interpret the phenomenon is introduced.
The summary and a brief discussion are given in \S $5$.

\section{Coupled Magnetic Systems}

We consider the composite mesoscopic spin-$\frac{1}{2}$ system, which 
is composed of a gapless chain and a gapped chain.
This attachment of two different mesoscopic magnetic systems 
means that heat transport between magnetic chains is possible by
spin-phonon interaction at the surface of contact between them. 
Here, it is assumed that the mean free 
paths in both magnetic systems are comparable to system size. 
These setups are depicted in Fig. $1$(a). 
We now use the following type of Hamiltonian to investigate the heat flow
at the nonequilibrium stationary state: 
\beq
{\cal H} &=& {\cal H}_{\rm L} + {\cal H}_{\rm R} + {\cal H}_{p}
+ {\cal H}'   \\
{\cal H}_{\rm p} &=& \sum_{k} \omega_{k} b_{k}^{\dagger} b_{k} 
\\
{\cal H}' &=& 
\sum_{k} \lambda_{{\rm L}k} s_{{\rm L},1}^{z} 
\lr b_{k} + b_{k}^{\dagger}
\rl 
+
\sum_{k} \lambda_{{\rm R}k} s_{{\rm R},1}^{z} \lr b_{k} 
+ b_{k}^{\dagger} \rl   
\eeq
Here ${\cal H}_{\alpha}$ $(\alpha = {\rm L} \,\, {\rm and}\,\, {\rm R})$ is 
the Hamiltonian of the left and right magnetic systems, and 
${\cal H}_{\rm p}$ expresses the phonon part. 
The operators $s_{{\rm L},1}^{z}$ and $s_{{\rm R},1}^{z}$ are 
the $z$-component of spins at the edges of the left and right 
magnetic systems, and they are assumed to interact with phonons 
localized at the contact between magnetic systems. 
These decompositions are schematically shown in Fig. $1$(b).
Here, the isotropic Heisenberg chain is assumed to be the gapless chain, 
but no concrete gapped chain is assumed at present.

We treat mesoscopic-scale systems so that 
mean free path is comparable to material size in both 
magnetic materials, where 
heat is conveyed without scattering within the material. 
In this case, the effective Hamiltonian is well 
described by the bosonization form of the spin systems.
Generally, bosonized Hamiltonians satisfy a ballistic 
feature of transport because 
the bosonized heat flow operator is commutable with the Hamiltonian
\cite{S03}.  
Thus, the bosonized form of the spin system automatically satisfies the 
characteristics of the mesoscopic nature of the magnetic systems. 
We adopt the following bosonized Hamiltonian:
\beq
{\cal H}_{\rm L} &=& {u\over 2} 
\int_{x_{0}}^{\infty} dx \, \left[ \Pi_{\rm L}^{2} (x) + 
\lr \partial_{x} \Phi_{\rm L} (x)\rl^{2} 
\right] \\
{\cal H}_{\rm R} &=& {u\over 2} 
\int_{x_{0}}^{\infty} dx \, \left[ \Pi_{\rm R}^{2} (x) + 
\lr \partial_{x} \Phi_{\rm R} (x)\rl^{2} + g_0 \Phi_{\rm R}^2 (x)
\right] \\
{\cal H}' &=& \sum_{\alpha={\rm L},{\rm R}}
\lr {1\over\sqrt{\pi}}\partial_{x}\Phi_{\alpha }(x_0) + 
\sum_{k} {\lambda_{\alpha k}  \over 2 }
\lr b_{k} + b_{k}^{\dagger}
\rl \rl^2   \label{hprime}
\eeq
The boson operators 
$\Phi_{\alpha} (x)$ and $\Pi_{\alpha'} (x')$ satisfy the
commutation relation $[ \Phi_{\alpha}(x) , \Pi_{\alpha'}(x' ) ]
=\ii \delta_{\alpha,\alpha'}\delta (x - x' )$. 
The operator ${1\over\sqrt{\pi}}\partial_{x}\Phi_{\alpha }(x_0)$
is the $z$-component of spin at the position $x_0$ in the $\alpha$th
spin chain.
In the interaction Hamiltonian ${\cal H}'$, the quadratic terms of
bosons are added only to avoid an unbounded state.
For simplicity of calculation, we adopt the bilinear form 
for the gapped Hamiltonian, which can be regarded as the 
approximated form of the sine-Gordon Hamiltonian 
${\cal H}_{\rm R}$ \cite{text,NF80}.
The sine-Gordon Hamiltonian can be mapped from many gapped magnetic chains.
The present gapped Hamiltonian gives the dispersion relation with 
the energy gap $\Delta = u \sqrt{g_0 } $.
We believe that this simplification does not cause different 
qualitative results because correct excitation spectrum and 
excitation eigenfunctions are not necessary for the phenomenon 
described in this paper.
\section{Heat Flow}

In this section, we calculate the heat flow at the steady state.
As explained in the previous section, the mesoscopic nature of
current is represented by the ballistic heat flow in the bosonized 
Hamiltonian. The next task is the realization of the nonequilibrium steady 
state of the system.
For simplicity of calculation, we assume an infinite size for 
the left and right magnetic systems, instead of connecting thermal reservoirs 
at the edges of the magnetic systems. By taking the infinite size, 
we can realize a nonequilibrium steady state as follows (see Fig. $1$(b)). 
We assume that both magnetic systems are initially separated 
and are in equilibrium with the temperatures 
$T_{\rm L}$ and $T_{\rm R}$.
Next, these systems are adiabatically connected with the phonon
system, and a steady state is realized in the long time limit.
The steady state can be analyzed by the Keldysh Green function
technique \cite{Keldysh}. 
To this end, the following operators are defined
\beq
X_{\alpha} &=& \sum_{k }\lambda_{\alpha k} 
\left( b_{\alpha k} + b_{\alpha k}^{\dagger } \right) , 
 \quad
Y_{\alpha} = \ii\sum_{k }\lambda_{\alpha k} 
\left( b_{\alpha k} - b_{\alpha k}^{\dagger } \right) 
\quad 
{\alpha}={\rm R} \,\,{\rm or}\,\, {\rm L}
\eeq
We further assume the simple form of coupling for spin-phonon interaction
as $\lambda_{{\rm L}k}=\lambda_{{\rm R}k}=\lambda_{k}$. We
do not discriminate between the indices 
$\alpha$ of operators $X_{\alpha}$ and 
$Y_{\alpha}$, and simply write $X$ and $Y$, respectively.
The heat flow is expressed by the Keldysh lesser Green function 
\beq
\lan J \ran &=& 
\langle {\ii\over \sqrt{\pi}} \partial_{x}\Phi_{1}(x_0 ,t) Y(t) \rangle 
= {\ii\over \sqrt{\pi}} 
G_{Y,\partial\Phi_{1}}^{<}(t , t) \nonumber \\
&=&-{1\over 2 \pi^{2}} 
 \int_{-\infty}^{\infty} d\Omega \,\,
{\rm Im}\left\{ 
G_{Y,X}^{r} (\Omega ) g_{\partial \Phi_{\rm L} ,\partial \Phi_{\rm L} }^{<} 
(\Omega )
+
G_{Y,X}^{<} (\Omega ) g_{\partial \Phi_{\rm L} ,\partial \Phi_{\rm L} }^{a}
(\Omega ) \right\}   \label{formula}
\eeq
Throughout this paper, we set $\hbar$ to be unity. 
Here $G_{Y,X}^{r} (\Omega )$ and $G_{Y,X}^{<} (\Omega )$ are
the Fourier transforms of the retarded and lesser Green functions:
\beq
G_{Y,X}^{r} (t,t' ) &=& -\ii\Theta (t-t')
\langle \left[ Y(t) , X(t' )\right]\rangle \nonumber \\
G_{Y,X}^{<} (t,t' ) &=&  -\ii \langle X(t') Y(t)\rangle \nonumber
\eeq
The functions $g_{A,B}^{<}(\Omega)$ and $g_{A,B}^{r,a}(\Omega)$ are
the Fourier transforms of the nonperturbed Green function:
\beq
g_{A,B}^{<}(t,t') &=&-\ii\langle B(t' ) A(t) \rangle_{0} \nonumber \\
g_{A,B}^{r,a}(t,t') &=& \mp \ii\Theta \lr \pm (t - t' )\rl
\langle [A(t) ,B(t' ) ]\rangle_{0} \nonumber
\eeq
Here $\langle ... \rangle_{0}$
denotes an average over a nonperturbed Hamiltonian.
The nonperturbed Green functions are readily calculated as
\beq
\left(
\begin{array}{cc}
g^{r,a}_{X_{k} X_{k} } (\Omega ) , & 
g^{r,a}_{X_{k} Y_{k} } (\Omega )  \\
g^{r,a}_{Y_{k} X_{k} } (\Omega ) , &
g^{r,a}_{Y_{k} Y_{k} } (\Omega )  
\end{array}
\right) 
&=& {\lambda^{2}_{k}/\pi \over \lr \Omega \pm \ii \varepsilon \rl^2 -
\omega_k }
\left(
\begin{array}{cc}
\omega_k  & - \ii \lr \Omega \pm i \varepsilon \rl \\
\ii \lr \Omega \pm \ii \varepsilon \rl & \omega_k
\end{array}
\right)  ,\\
\ii g_{\partial\Phi_{\alpha} ,\partial\Phi_{\alpha} }^{<} (\Omega ) 
&=& 
\left\{ 
\begin{array}{cl}
2 {\sqrt{\Omega^2 - \Delta_{\alpha}^{2}} \over v^2 } {{\rm sgn}(\Omega )
\over \ee^{\beta\Omega} -1 } 
& \qquad {\rm for} 
\quad |\Omega| > \Delta_{\alpha} \\
0 &  \qquad {\rm for} \quad
|\Omega| \le  \Delta_{\alpha} 
\end{array}
\right. ,\\
g_{\partial \Phi_{\alpha} , \partial \Phi_{\alpha}}^{r,a} (\Omega )
&=& 
\left\{
\begin{array}{cc}
\mp
{\ii\over v^2}{\rm sgn} (\Omega ) \sqrt{ \Omega^2 - \Delta_{\alpha}^2 } 
& \qquad {\rm for} \quad | \Omega | > \Delta_{\alpha}  \\
{\sqrt{ \Delta_{\alpha}^2 -\Omega^2 } \over v^2} 
& \qquad {\rm for} \quad  |\Omega | \le  \Delta_{\alpha} 
\end{array}
\right. ,
\eeq
where $\alpha$ is L or R, and $\Delta_{\rm L}= 0$, 
and $\Delta_{\rm R} = u \sqrt{g_0 }$. Here, we neglected the effects of
the quadratic terms of the bosons in the Hamiltonian (\ref{hprime}), 
which is not expected to have a large contribution to the
stationary state.

In realistic systems, it is natural to expect that the phonon system
at the contact between magnetic systems becomes a local 
equilibrium state at the steady state.  
To realize the thermalization of the phonon part, we adopt
the so-called self-consistent reservoir (SCR) by connecting 
an extra infinite system, which can be regarded as the reservoir with 
some temperature $T_{\rm S}$ at the initial time.
The temperature $T_{\rm S}$ is determined under the
condition that heat flow vanishes between the reservoir and 
the phonon system 
at the stationary state as depicted in Fig. $2$(a). 
Such an extra system induces the thermalization of phonons, and the degree
of a local equilibrium state is controlled by the coupling strength
between the SCR and the phonon system \cite{BRV70}. 
Since this extra system is simply adopted only to induce the thermalization 
of the phonon system, we have many choices for the system.
For simplicity, we now use the spin system for the SCR, 
which is the same form of ${\cal H}_{\rm L}$ because
all Green functions are already known. 
The interaction Hamiltonian where the phonon couples with the SCR
is 
\beq
{\cal H}_{\rm SCR-P} &=& \lambda 
\lr {1\over\sqrt{\pi}}\partial_{x} \Phi_{\rm SCR}(x_0) + 
\sum_{k} {\lambda_{\alpha k}  \over 2 }\lr b_{k} + b_{k}^{\dagger}
\rl \rl^2 ,
\eeq
where ${1\over\sqrt{\pi}}\partial_{x} \Phi_{\rm SCR}(x_0) $ 
is the z-component of spin at $x_0$ in the SCR, and
$\lambda$ is the coupling strength between SCR and the phonon system.
The strength of $\lambda$ controls the degree of local equilibrium.
The spin system for the SCR is the same form as that ${\cal H}_{\rm L}$;
${\cal H}_{\rm SCR} = {u\over 2} 
\int_{0}^{\infty} dx \, \left[ \Pi_{\rm SCR}^{2} (x) + 
\lr \partial_{x} \Phi_{\rm SCR} (x)\rl^{2} 
\right] .$  
The Green functions $G_{Y,X}^{r} (\Omega )$ and $G_{Y,X}^{<} (\Omega )$ in 
the formula (\ref{formula}) are calculated by
the Keldysh relation \cite{Keldysh} with
the nonperturbed Green functions of 
the phonon system, two magnetic systems and the system of SCR. 

Let us now see the properties of heat flow in these setups.
Two cases are investigated, i.e., $(T_{\rm L}, T_{\rm R})=(\Delta T, 0)$
and $(T_{\rm L}, T_{\rm R})=(0, \Delta T )$. The heat flows in
the former and the latter cases are written as 
$J_{\rm L}$ and $J_{\rm R}$, respectively. In these two cases, 
the behavior of heat flow is investigated for various coupling strengths
$\lambda$. 
Numerical calculation is self-consistently performed to satisfy 
the condition of SCR. Here, we took $1000$ boson numbers for the phonon 
system with the ohmic spectral density. The energy gap is $\Delta=3.0$
in the unit of $u$.
The temperature dependences of heat flow are shown in Fig. $2$(b) for
various coupling strength $\lambda$. 
As shown in Fig. $2$(b), 
in the case of $\lambda=0$, no asymmetry of heat flow
is observed. However as $\lambda$ increases, asymmetry appears.
For a larger $\lambda$, the amplitude of heat flow decreases, which is 
due to the scattering effect of energy by the SCR. 
In the case of finite $\lambda$, $|J_{\rm R}|$ is always larger than 
$|J_{\rm L}|$.

\section{Extreme Model to understand Mechanisms}

As shown in the previous section, asymmetric current can occur
in the presence of the thermalization of phonons. This means that 
the phonon system has some temperature defined in the local equilibrium 
state. In this section, we consider a simple phenomenological 
explanation of the asymmetric heat flow regarding the phonon 
system as {\em the phonon reservoir}.

To this end, the following two problems are considered.
i) First, what amount of the stationary heat flow 
is obtained in one boson system when it is coupled to two different 
reservoirs with different temperatures ? 
Here the boson system is the main system of interest.
ii) Second, the following situation is studied
using the general formula of heat flow derived in the first problem.
Suppose that the different two boson systems sandwich a phonon reservoir
with some temperature, $T_{\rm S}$. At the edges of the boson system, 
other phonon reservoirs are connected with the 
temperatures $T_{\rm L}$ and $T_{\rm R}$
(see the schematic explanation of the situation in the inset in the 
Fig. $3$). 
What is the characteristic of heat flow, 
when the temperature $T_{S}$ is determined under the condition that 
both stationary heat flows within the boson systems are equal ? 
This extreme situation actually explains the asymmetry of heat flow.

We study the first problem by considering the dynamics for the Hamiltonian;
\beq
{\cal H}_{\rm tot} &=& {\cal H} + 
\sum_{\alpha=1,2} \nu
\left( X_{\alpha} - \sum_{k} \gamma_{\alpha,k} 
(a_{\alpha,k} + a_{\alpha,k}^{\dagger} ) \right)^2 + \sum_{\alpha,k}
\omega_{k} a_{\alpha,k}^{\dagger} a_{\alpha,k} , \\
{\cal H} &=& \sum_{k} \Omega_{k} b_{k}^{\dagger} b_{k} , \label{hamham}\\
X_{\alpha} &=& \sum_{k} \mu_{\alpha,k} 
\left( b_{k}^{\dagger}  + b_{k} \right) ,
\eeq 
where ${\cal H}$ is the boson system of interest and 
the operators $a_{\alpha ,k}$ and $a_{\alpha ,k}^{\dagger}$ are 
the annihilation and creation phonon operators in the phonon reservoirs.
The operators $X_{1}$ and $X_{2}$ are the 
system's operators attached to
the phonon reservoirs. We assume that 
the phonon reservoirs have spectral densities $I_{\alpha} (\omega )$
and temperatures $T_{\alpha}$. 

By standard projection operator
technique \cite{KTH}, we obtain the Redfield-type master equation for a 
reduced 
density operator \cite{R65,stm00} up to the second order of $\nu$
\beq
{\partial \rho (t) \over \partial t }
&=& -\ii \left[ {\cal H } , \rho (t) \right] -
\nu^2 {\cal L}_{1}\rho (t) 
- 
\nu^2 {\cal L}_{2}\rho (t)  \\
{\cal L}_{\alpha } \rho(t) &=& 
\left[ X_{\alpha} , R_{\alpha} \rho (t) \right] + 
\left[ X_{\alpha} , R_{\alpha} \rho (t) \right]^{\dagger} 
 \\
R_{\alpha } & = & \sum_{k}  \mu_{k} 
I_{\alpha } ( \Omega_{k } ) n_{\alpha } (\Omega_{k }) 
\left( 
\ee^{ \beta_{\alpha} \Omega_{k} } b_{k } + b_{k }^{\dagger } 
\right) 
\eeq
Here, $n_{\alpha} ( \Omega_{k} )$ is the Bose distribution,
$n_{\alpha } ( \Omega_{k} ) = 1/(\ee^{\beta_{\alpha } 
\Omega_{k}} - 1 )$.
When we expand the stationary density matrix as 
$\rho_{\rm st} = \rho_{\rm st}^{(0)} + \nu^2 \rho_{\rm st}^{(1)} + \cdots$,
we obtain the zeroth order of density matrix as
\beq
\rho_{\rm st}^{(0)}  &=& \Pi_{k} 
{ \ee^{ - B_{k} \Omega_{k} b_{k}^{\dagger} b_{k }} \over Z } ,
\qquad 
B_{k }= {1\over \Omega_{k } } \ln
\left[ {\sum_{\alpha = {1},{2} }   
\mu_{\alpha , k}^{2}  I_{\alpha} (\Omega_{k} ) n_{\alpha }(\Omega_{k } ) 
\ee^{\beta_{\alpha} \Omega_{k }} \over 
\sum_{\alpha = {1},{2} } \mu_{\alpha, k}^{2} 
I_{\alpha }  (\Omega_{k} ) n_{\alpha } (\Omega_{k } )} 
\right] .
\eeq

The heat flow operator from the first reservoir to system 
is derived from the continuity equation of energy 
\beq
{\hat j} = 
\nu^2 {\cal H} {\cal L}_{1}.
\eeq
Thus, using the zeroth order stationary density matrix $\rho_{\rm st}^{(0)}$, 
we obtain the leading order of average current 
$\langle j \rangle= \nu^2 {\rm Tr} \left( 
{\cal H} {\cal L}_{1} \rho_{\rm st}^{(0)}\right)$. 
The average current $\langle j \rangle$ is readily calculated,
\beq
\langle j \rangle &=&
\nu^2 \int d\varepsilon {\cal D} (\varepsilon ) \varepsilon 
\left[ 
{
\mu_{1} (\varepsilon ) I_{1} (\varepsilon )
\mu_{2} (\varepsilon ) I_{2} (\varepsilon )
\over
\mu_{1} (\varepsilon ) I_{1} (\varepsilon )
+
\mu_{2} (\varepsilon ) I_{2} (\varepsilon )
}
\right] \lr n_{2}(\varepsilon ) - n_{1} (\varepsilon )\rl , \label{cf}
\eeq    
where ${\cal D} (\varepsilon )$ is the density of states at the energy
$\varepsilon$, and $\mu_{\alpha} (\varepsilon) 
= \sum_{k} \mu_{\alpha ,k }^2 \delta \lr \varepsilon - \Omega_{k} \rl$.
The function of $\mu_{\alpha, k}$ depends on detailed systems. 
Below we consider the special 
case with $\mu_{1,k}^{2}=\mu_{2,k}^{2}$ for simplicity.
This case is actually realized in some boson systems \cite{consider}. 
In this case, eq.(\ref{cf})
becomes simpler because of $\mu_1 (\varepsilon) = \mu_2 (\varepsilon)$.

We are now in the position to discuss the second problem and clarify 
the essential mechanism of asymmetric heat flow. Suppose that the 
different boson systems are connected via a phonon reservoir with 
temperature $T_{\rm S}$, and are also attached to phonon reservoirs with 
different temperatures, $T_{\rm L}$ and $T_{\rm R}$ (see the inset in Fig. 3).
We call the heat flow within the left and right boson systems as 
$j(T_{\rm L}, T_{\rm S})$ and $j(T_{\rm S}, T_{\rm R})$, respectively. 
At the stationary state, it is demanded that both currents are equal, i.e.,
\beq
j(T_{\rm L}, T_{\rm S}) = j(T_{\rm S}, T_{\rm R}),
\eeq
from which the temperature $T_{\rm S}$ is self-consistently determined. 
Note that the heat flow is measured under these conditions.

We now demonstrate that the asymmetry of current actually occurs when
the energy densities of the left and right systems, 
${\cal D}_{\rm L}(\varepsilon )$
and ${\cal D}_{\rm R}(\varepsilon )$ are different.
We consider the simplest case of ${\cal D}_{\rm L} (\varepsilon ) = 
d_{0}\varepsilon $ and ${\cal D}_{\rm R} (\varepsilon ) = 
d_{0}\sqrt{\varepsilon^2 + \Delta^2}$. 
In this case, we calculate the energy flow $J_{\rm L}$ for the temperature set
$(T_{\rm L},T_{\rm R})=(\Delta T ,0)$, and
$J_{\rm R}$ for $(T_{\rm L},T_{\rm R})=(0, \Delta T )$.
In Fig. $3$, we present the results of $J_{\rm R}$ and $J_{\rm L}$ 
under these conditions. Here, energy gap $\Delta=3.0$, and 
we used the same phonon reservoir with ohmic spectral density. 
As shown, we find that the asymmetry appears in the heat flow, 
and the qualitative results in the previous section are 
reproduced.

\section{Summary}

We discussed the possibility of observation of asymmetric heat flow
in a coupled magnetic material. We particularly focused on 
mesoscopic-scale magnetic systems, where mean free path 
is comparable to system size. 
We found that an asymmetric current is observed 
as a result of the thermalization of phonons at the contact 
between magnetic materials. The present experimental setup is realistic,
and we have many choices of realistic magnetic materials.
The asymmetry of heat flow is a key ingredient for heat pumping. 
It will be important to discuss a realistic setup of heat pumping and 
the efficiency \cite{sn}, although we do not discuss it here.

In a gapped spin chain, a phononic contribution may appear in realistic 
experiments. Actually in CuGeO$_{3}$, which shows spin-Peierls transition, 
there exists a crossover from magnetic heat transport to phononic 
heat transport 
when the temperature decreases across the spin-Peierls temperature
\cite{ATSDTFK98}. 
The quantitative properties depend on spin-phonon interaction in the bulk
materials. Thus, to realize spin heat transport in all temperature 
regimes, we must find a suitable material with very weak spin-phonon 
interaction.

In the present study, we consider  
small magnetic systems, where the analysis becomes easy.
In the case that mean free path is much smaller than 
material size, treatment is much more difficult than the present 
analysis because most gapped magnetic systems are nonintegrable systems,
and use of the bosonized Hamiltonian may not be justified\cite{HHCB02}.
It will also be important to study such a region,
because it is not clear whether observation of asymmetric heat flow 
requires the ballistic properties of current.
To answer this question, real experiments would be very useful. 

\newpage 
\begin{figure}[tb]
\begin{center}
\includegraphics[scale=0.5,clip]{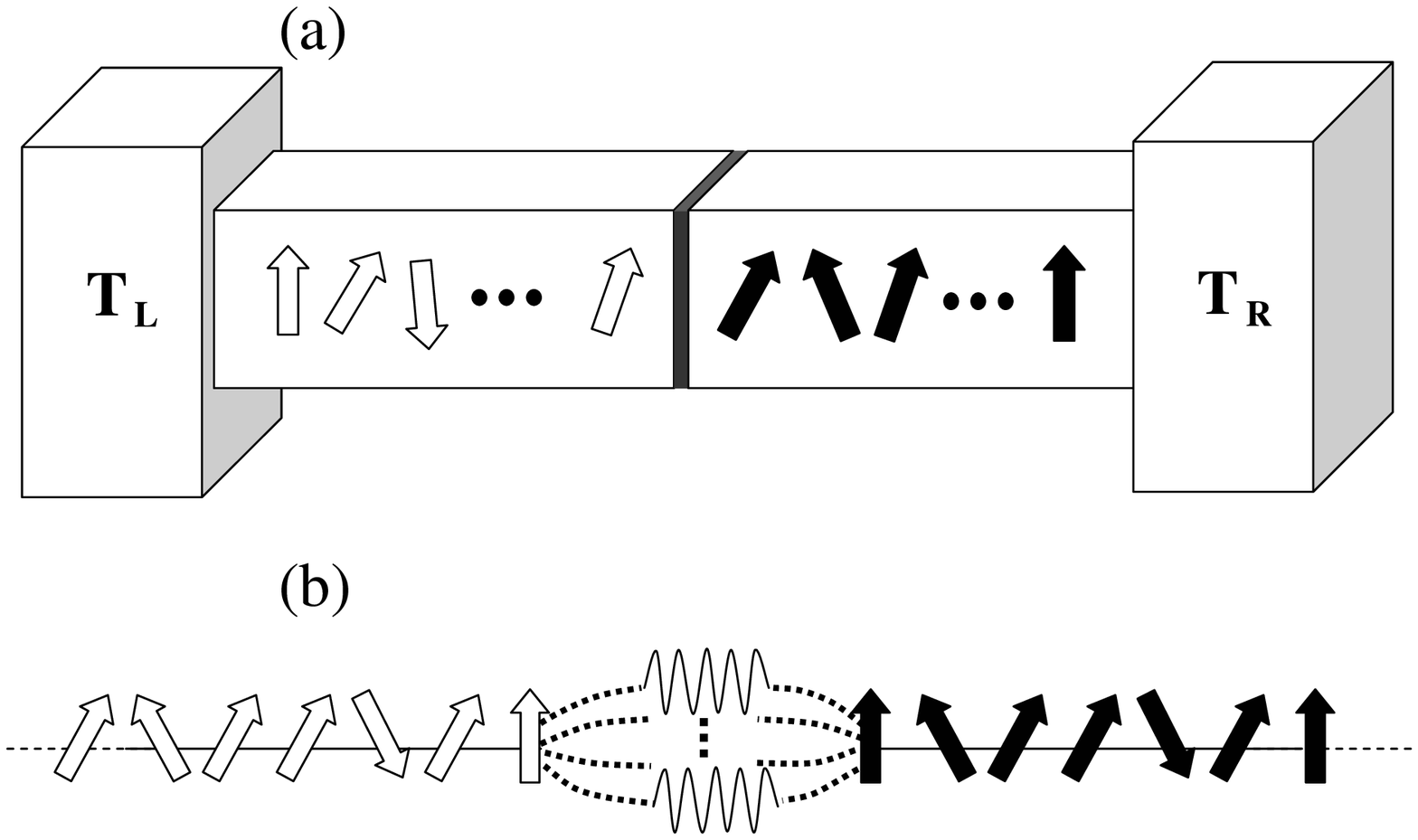}
\end{center}
\vspace*{-3cm}
\caption{(a): Experimental setup of coupled magnetic system. 
It is assumed that no magnetic interactions exist 
between the two magnetic systems. The heat can be conveyed by phonons 
at the surface of the contact. 
(b): Theoretical setup. At the contact between the magnetic systems, 
spin-phonon interaction exists.}
\label{fig.1}
\end{figure}
\begin{center}
Fig.1, Keiji Saito
\end{center}

\newpage
\begin{figure}[tb]
\begin{center}
\includegraphics[scale=0.5,clip]{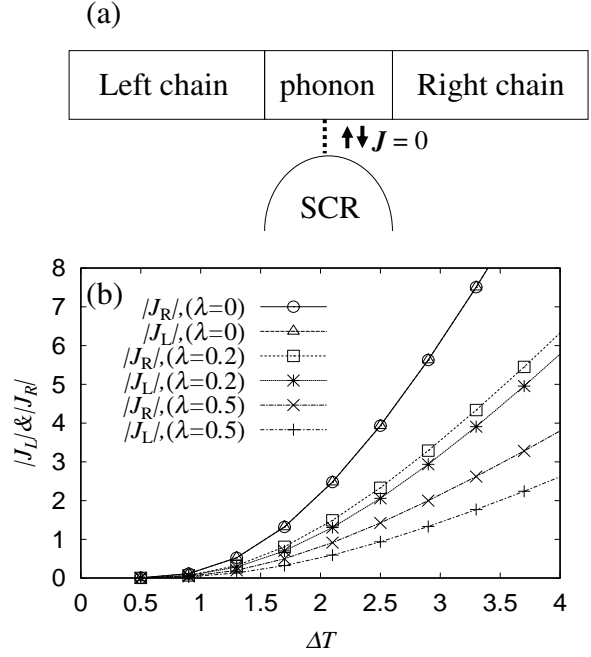}
\end{center}
\caption{(a) Schematic explanation of structure of self-consistent reservoir. Between the SCR and the phonon, there is no heat flow at the stationary
state. Thermalization occurs at the phonon part with a finite coupling 
strength, $\lambda$.   
(b) $|J_{\rm L}|$ and $|J_{\rm R}|$ as functions of $\Delta T$ for 
various coupling strengths $\lambda$ of SCR. 
Here, the energy gap of the right spin chain is $3.0$ in the unit of $u$.
The unit of the heat flow is $u^2/\hbar$.}
\label{fig.2}
\end{figure}
\begin{center}
Fig.2, Keiji Saito
\end{center}

\newpage 
\begin{figure}[tb]
\begin{center}
\includegraphics[scale=0.5,clip]{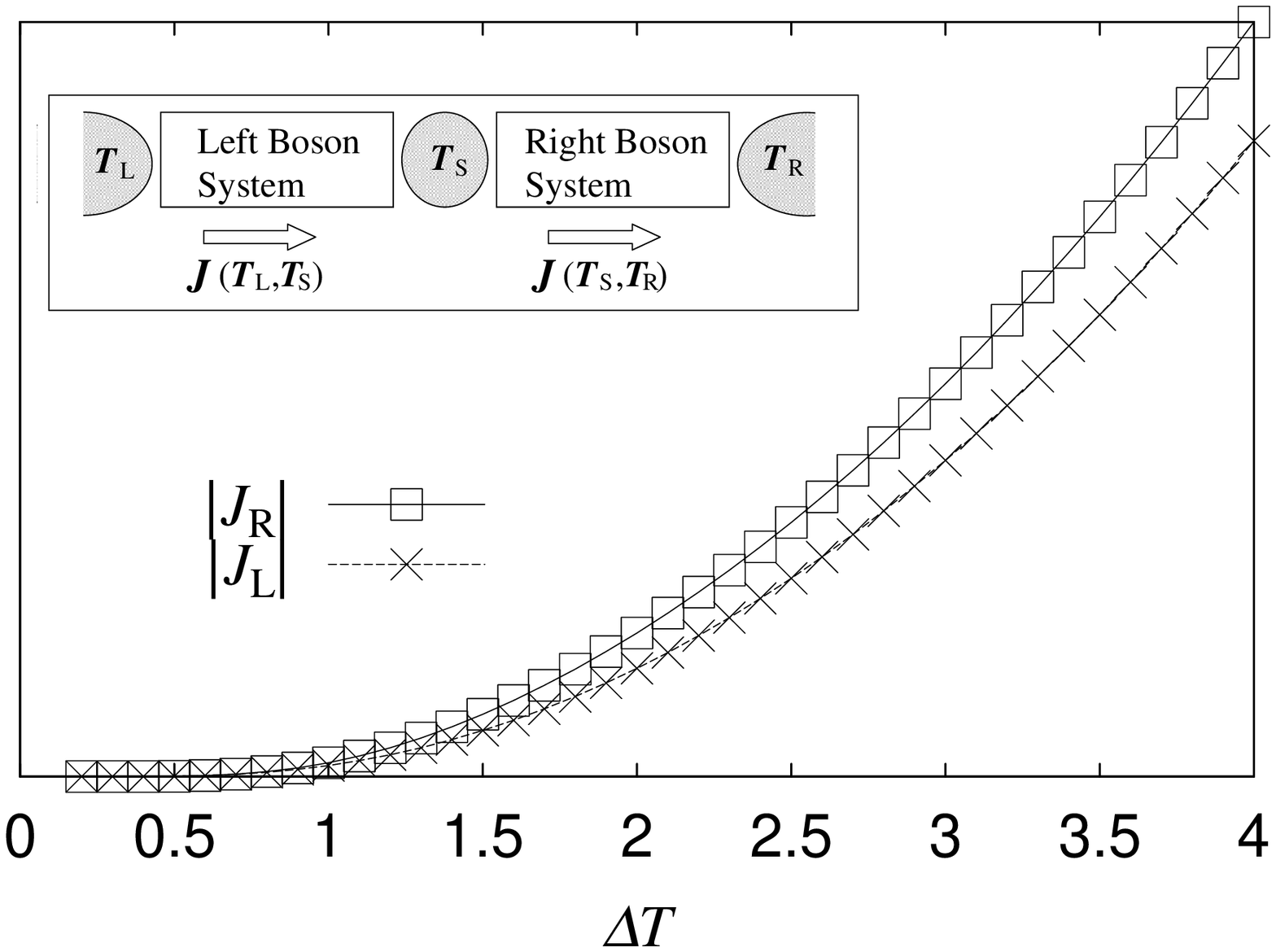}
\end{center}
\caption{Simple explanation of asymmetric heat flow 
using phonon reservoirs with  
temperatures $T_{\rm L}$, $T_{\rm S}$, and $T_{\rm R}$.
The temperature 
$T_{\rm S}$ is determined by the condition $J(T_{\rm L}, T_{\rm S})
=J(T_{\rm S}, T_{\rm R})$}
\label{fig.3}
\end{figure}
\begin{center}
Fig.3, Keiji Saito
\end{center}


\begin{thebibliography}{99} 
\bibitem{SFGOVR00} 
A. V. Sologubenko, K. Gioann$\grave{\rm o}$, H. R. Ott,
	A. Vietkine and A. Revcolevschi: Phys. Rev. B \textbf{62} (2000) 
R6108.
\bibitem{SFGOVR00b} 
A. V. Sologubenko, K. Gioann$\grave{\rm o}$, H. R. Ott,
	A. Vietkine and A. Revcolevschi: Phys. Rev. B \textbf {64} (2000) 
054412.
\bibitem{ATSDTFK98} Y. Ando, J. Takeya, D. L. Sisson, S. G. Doettinger, 
I. Tanaka, R. S. Feigelson and Kapitulnik: Phys. Rev. B \textbf{58} (1998) R2913.
\bibitem{HLUKZFKU01} M. Hofmann et al: Phys. Rev. Lett. \textbf{87}
(2001) 047202-1.
\bibitem{SGOARBT00}
A.V. Sologubenko, K. Gioann$\grave{\rm o}$, H. R. Ott, U. Ammerahl,
	A. Revcolevschi, D. F. Brewer and A. L. Thomson: Physica B 
\textbf{284-288} (2000) 1595.
\bibitem{SGOAR00} 
A.V. Sologubenko, K. Gioann$\grave{\rm o}$, H. R. Ott, U. Ammerahl and
	A. Revcolevschi: Phys. Rev. Lett. \textbf{84} (2000) 2714.
\bibitem{KINAKMTK01}
K. Kudo, S. Ishikawa, T. Noji, T. Adachi, Y. Koike, K. Maki, S. Tsuji and
K. Kumagai: J. Phys. Soc. Jpn. \textbf{70} (2001) 437.
\bibitem{ZNP97}
X. Zotos, F. Naef and P. Prelovsek: Phys. Rev. B \textbf{55} (1997) 11029.
\bibitem{STM96}
K. Saito, S. Takesue and S. Miyashita: Phys. Rev. E \textbf{54} (1996) 
2404.
\bibitem{theo1}
J. V. Alvarez and C. Gros: Phys. Rev. Lett. \textbf{89} (2002) 156603.
\bibitem{theo2}
J. V. Alvarez and C. Gros: Phys. Rev. Lett. \textbf{88} (2002) 077203.
\bibitem{theo3}
S. Fujimoto and N. Kawakami: Phys. Rev. Lett. \textbf{90} (2003) 197202.
\bibitem{theo4}
E. Orignac, R. Chitra and R. Citro: Phys. Rev. B \textbf{67} (2003) 
134426.
\bibitem{theo5}
E. Shimshoni, N. Andrei and A. Rosch: Phys. Rev. B \textbf{68} 
(2003) 104401.
\bibitem{theo6}
X. Zotos: Phys. Rev. Lett. \textbf{92} (2004) 067202.
\bibitem{theo7}
C. Hess P. Ribeiro, B. Buchner, H. ElHaes, G. Roth, U. Ammerahl 
and A. Revcolevschi: cond-mat/0506595.
\bibitem{theo8}
P. Jung, R. W. Helmes and A. Rosch: cond-mat/0509615.
\bibitem{HHCB02}
F. Heidrich-Meisner, A. Honecker, D. C. Cabra and W. Brenig: Phys. Rev. 
B \textbf{66}, (2002) 140406(R).
\bibitem{S03}K. Saito: Phys. Rev. B \textbf{67} (2003) 064410.
\bibitem{TPC02}
M. Terraneo, M. Peyrard and G. Casati: Phys. Rev. Lett. \textbf{88}
(2002) 094302. 
\bibitem{LWC04}
B. Li, L. Wang and G. Casati: Phys. Rev. Lett. \textbf{93} (2004)
184301.
\bibitem{SN1}
D. Segal and A. Nitzan: Phys. Rev. Lett. \textbf{94} (2005) 034301.
\bibitem{Keldysh}
L. V. Keldysh, Zh. \'{E}ksp. Theor. Fiz. \textbf{47} (1964) 1515
[Sov. Phys. JETP \textbf{20} (1965) 1018.].
\bibitem{text}
A.O. Gogolin, A.A. Nersesyan and A.M. Tsvelik: 
Bosonization and Strongly Correlated Systems 
(Cambridge University Press, 1998).
\bibitem{NF80}
T. Nakano and H. Fukuyama: J. Phys. Soc. Jpn, \textbf{49} (1980) 1679.
\bibitem{BRV70}
M. Bolsterli, M. Rich and W. M. Visscher: Phys. Rev. A \textbf{1}
 (1970) 1086.
\bibitem{KTH}
R. Kubo, M. Toda and N. Hashitsume: Statistical Physics II
(Springer-Verlag, New York, 1985).
\bibitem{R65}
A. G. Redfield: Adv. Magn. Reson. \textbf{1} (1965) 1.
\bibitem{stm00}
K. Saito, S. Takesue and S. Miyashita: Phys. Rev. E \textbf{61} (2000)
2397.
\bibitem{consider}
For example, in case of the quantum harmonic chain whose 
Hamiltonian:
${\cal H} = \sum_{n=1}^{N}{p_{n}^{2} \over 2m} + 
\sum_{n=0}^{N} {m\omega_{0}^{2}\over 2} (x_{n+1}-x_{n})^2 
\,\, (x_{0}=x_{N+1}=0)$, $X_{1}$ and $X_{2}$ are
$x_1$ and $x_{N}$. The functions $\mu_{1,k}$ and $\mu_{2,k}$ 
are 
$\mu_{1,k}=\sqrt{{\hbar\over 2(N+1)m\omega_{0}}} 
{\sin k \over \sqrt{\sin (k/2)}}  $, and 
$\mu_{2,k}=\sqrt{{\hbar\over 2(N+1)m\omega_{0}}} 
{\sin (Nk) \over \sqrt{\sin (k/2)}}$,
where $k= \pi \ell / (N+1) \,\, (\ell = 1,2,\cdots , N)$. These
functions satisfy $\mu_{1,k}^{2}=\mu_{2,k}^{2}$. 

\bibitem{sn}
D. Segal and A. Nitzan, cond-mat/0510262/.
\end{thebibliography}
\end{document}